\documentclass[prd,notitlepage,showpacs,letterpaper,showpacs,nofootinbib]{revtex4-1}
\usepackage{graphicx,amsmath,amssymb,amsfonts,latexsym,color,dcolumn,bm,subfigure,ulem}

\usepackage{verbatim}

\usepackage{hyperref}

\begin{document}

\title{Many-Body Neutrino-Exchange Interactions and Neutrino Mass: \\ Comment on Phys. Rev. Lett. {\bf 120}, 223202 (2018)}
\author{E. Fischbach}
\affiliation{Department of Physics  and Astronomy, Purdue University, West Lafayette, Indiana 47907, USA}
%\author{D. E. Krause}
%\affiliation{Physics Department, Wabash College, Crawfordsville, Indiana 47933, USA}
%\affiliation{Department of Physics and Astronomy, Purdue University, West Lafayette, Indiana 47907, USA}

\date{ \today}

%\begin{abstract}
%Abstract goes here
%\end{abstract}

\maketitle

This Comment corrects an erroneous remark by Stadnik contained in Ref.~\cite{Stadnik} to the effect that many-body neutrino-mediated forces are suppressed in all types of stars.  As we discuss below, such forces were considered in great detail in Refs.~\cite{Fischbach AoP,Woodahl} and shown to produce catastrophically large energy densities in neutron stars and white dwarfs, unless neutrinos had a minimum mass $m \gtrsim 0.4$~eV/$c^{2}$.  Stadnik attributes his remark to an unpublished preprint \cite{Smirnov}.

Many-body effects arising from neutrino-exchange had been considered prior to Ref.~\cite{Fischbach AoP} by other authors including Feynman \cite{Feynman} and Hartle \cite{Hartle}.  The possibility of catastrophically large effects arising from many-body forces is suggested by the work of Primakoff and Holstein~\cite{Primakoff} dealing with many-body effects in atomic and nuclear systems.  These authors noted that in a nucleus containing $N$ particles the magnitude of the total $k$-body interaction ($k = 2, 3, 4, \ldots$) grows as the binomial coefficient
\begin{equation}
{N \choose k}= \frac{N!}{k! (N-k)!}.
\label{binomial}
\end{equation}
For a typical neutron star with $N \simeq 10^{57}$, combinatorics can thus lead to catastrophically large effects, as shown in detail in Refs.~\cite{Fischbach AoP} and \cite{Woodahl}.

To understand the arguments in Ref.~\cite{Smirnov} concerning Pauli-suppression of many-body neutrino exchange forces, we consider as an example the Hulse-Taylor binary system PSR 1913+16.  The mass $M_{1}$ of this pulsar is known to be $M_{1} = 1.4411M_{\odot}$, so that the number $N$ of neutrons is then approximately $N = 1.7 \times 10^{57}$, and we can take for its radius $R \simeq 10~{\rm km} \equiv R_{10}$.  The relevant parameter determining the many-body neutrino-exchange contributions is then given in terms of the Fermi constant $G_{F}$ by Eq.~(5.34) in Ref.~\cite{Fischbach AoP},
\begin{equation}
\eta(N,R_{10}) \equiv \frac{(G_{F}/\hbar c)N}{R_{10}^{2}} = 7.6 \times 10^{12}.
\label{parameter}
\end{equation}
It follows from Eq.~(\ref{parameter}) that in a neutron star for any spherical volume or sub-volume of radius $R$, the energy grows as $7.6 \times 10^{6}(R/1~{\rm cm})$.  As shown in Ref.~\cite{Fischbach AoP} for the Hulse-Taylor pulsar, the ratio of the many-body neutrino-exchange energy $W$ exceeds the total mass $M_{\rm U}$ of the known universe by the ratio
\begin{equation}
\frac{W}{M_{\rm U}} \simeq 10^{(2 \times 10^{59} -57 - 99)}.
\label{M U ratio}
\end{equation}

It is clear from the preceding discussion that any mechanism that could suppress the many-body neutrino-exchange contribution in any neutron star (or in any sub-volume of a neutron star) to the level required by Eqs.~(\ref{parameter}) and (\ref{M U ratio}) must be of a universal nature, such as minimum value of the neutrino mass, rather than one depending on the detailed characteristics of individual neutron stars and white dwarfs.

\begin{figure}[t]
\includegraphics[height=2.4in]{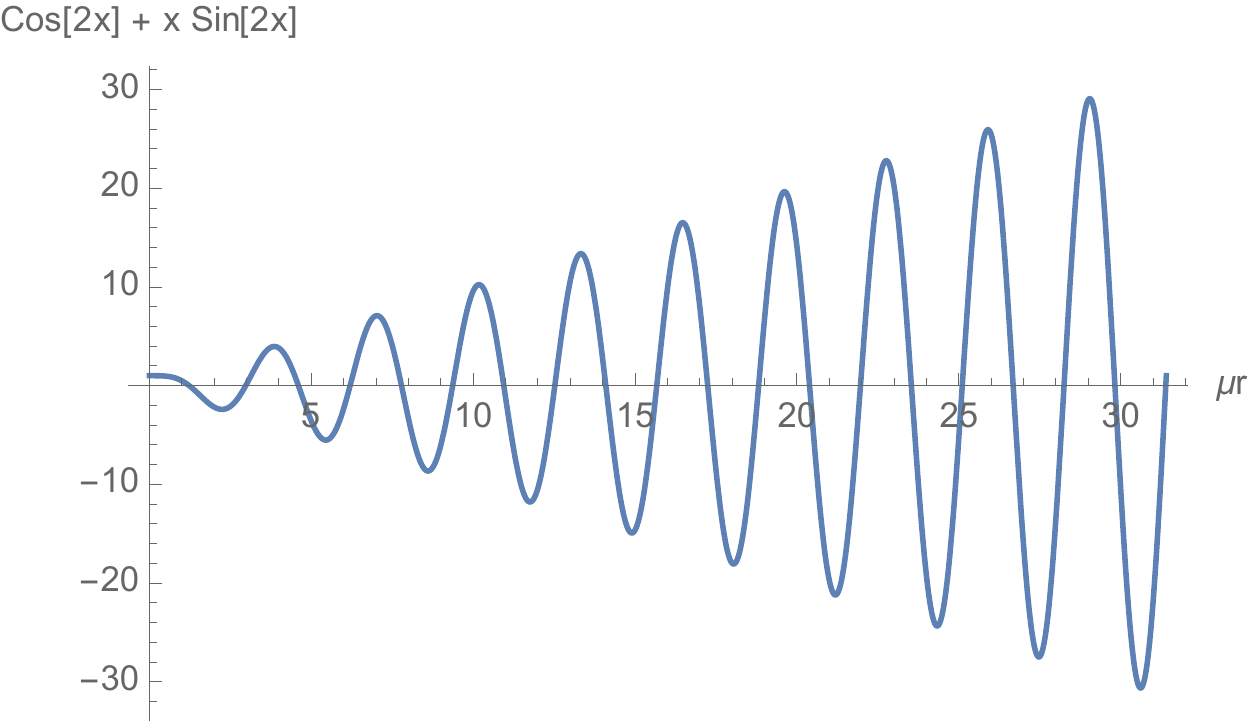}
%\resizebox{\columnwidth}{!}{\includegraphics{VMPlot}}  
  \caption{Plot of the oscillatory portion of $V_{\mu}^{(2)}(r)$ given by Eq.~(\ref{V mu}).}
  \label{V mu plot}
\end{figure}

 This conclusion is already evident from Eq.~(14) of Ref.~\cite{Woodahl} who exhibit the functional form of the simple 2-body exchange in the presence of Pauli-blocking,
\begin{equation}
V_{\mu}^{(2)}(r) = \frac{(G_{F}a_{f})^{2}}{4\pi^{3}r^{5}}\left[\cos(2\mu r) + \mu r\sin(2\mu r)\right],
\label{V mu}
\end{equation}
where $\mu$ is the local chemical potential.  A plot of the oscillatory factor in Eq.~(\ref{V mu}) as a function of $x = \mu r$ is shown in Fig.~\ref{V mu plot}.  
The authors of Ref.~\cite{Smirnov} note that $\cos(2\mu r)$ and $\sin(2\mu r)$ can be rapidly oscillating factors.  However, we see from Fig.~\ref{V mu plot} that even if the 2-body oscillatory factor happened to accidentally vanish for a spherical volume of radius $r$ given a particular value of $\mu$, any other volume with radius $r(1 + \epsilon)$, $\epsilon \ll 1$, will again lead to a catastrophically large energy density in light of the results of Eqs.~(\ref{parameter}) and (\ref{M U ratio}).

A detailed examination of the 4-body neutrino-exchange contributions illustrates why the requisite cancellations suggested in Refs.~\cite{Stadnik} and \cite{Smirnov} are extremely unphysical.  Consider Eq.~(16) of Ref.~\cite{Woodahl} which gives the 4-body analog of Eq.~(\ref{V mu}):
\begin{equation}
\tilde{V}_{0}^{(4)} = \frac{(G_{F}a_{n})^{4}}{2\pi^{5}P_{4}S_{4}} \left[\cos(\mu S_{4})
	\left(\frac{3}{S_{4}^{4}} - \frac{3\mu^{2}}{2S_{4}^{2}} + \frac{\mu^{4}}{8}\right)
	+ \frac{\mu\sin(\mu S_{4})}{S_{4}}\left(\frac{3}{S_{4}^{2}}- \frac{\mu^{2}}{2}\right)\right],
\end{equation}
\begin{figure}[t]
\includegraphics[height=2.5in]{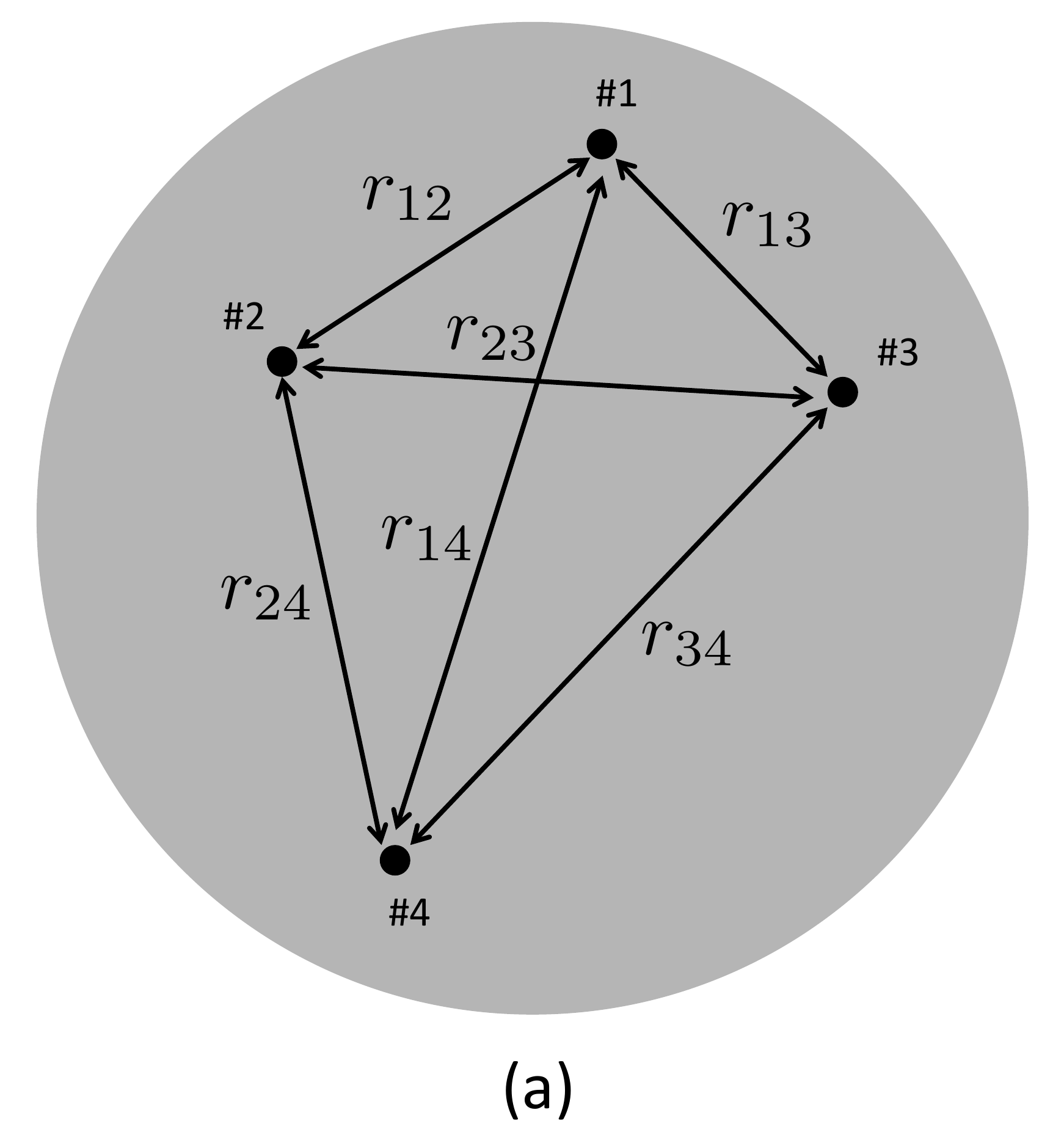}\includegraphics[height=2.5in]{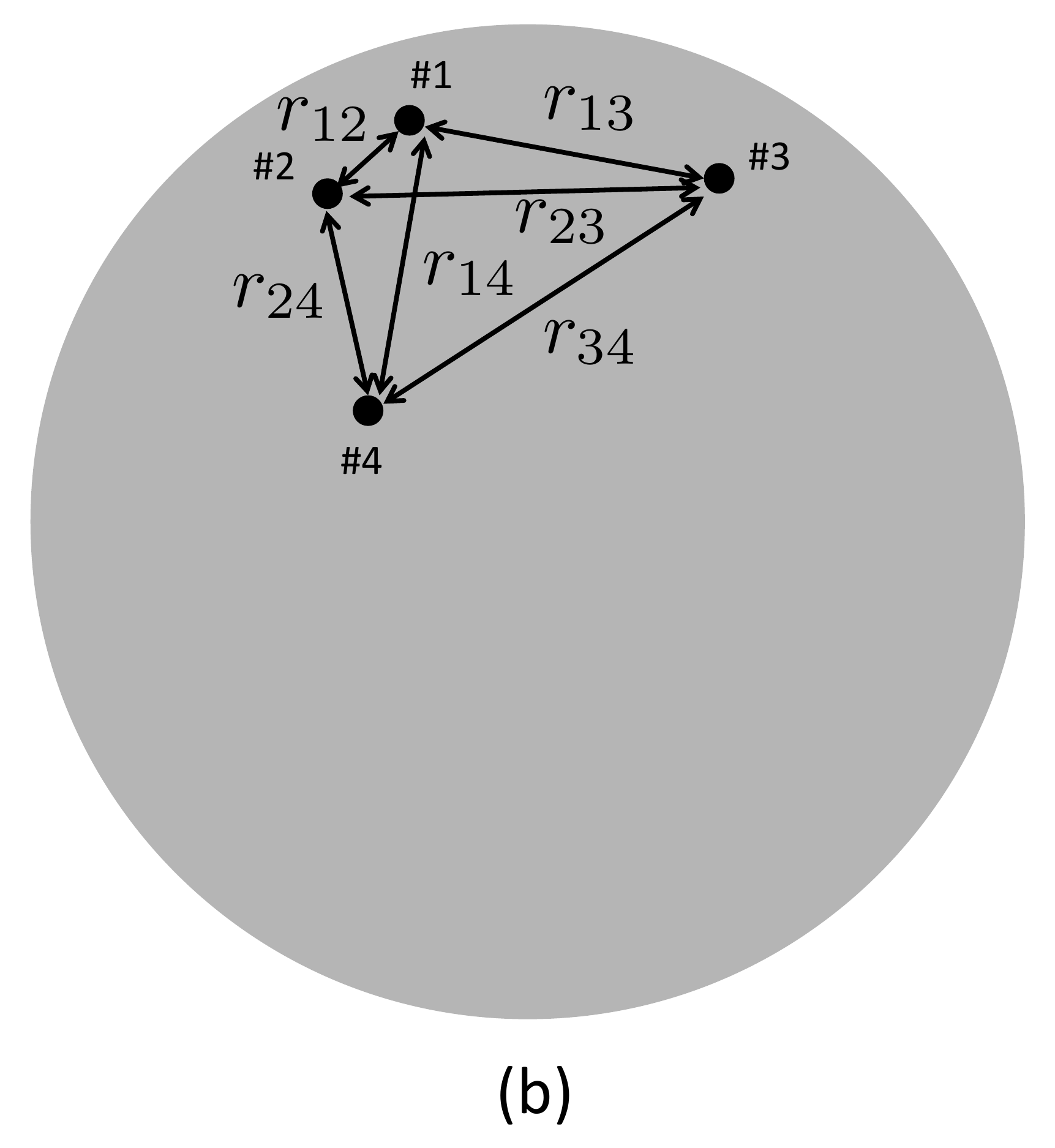}
%\resizebox{\columnwidth}{!}{\includegraphics{VMPlot}}  
  \caption{Two arbitrary configurations (a) and (b) of four neutrons in a neutron star showing that it is impossible that $S_{4}^{(a)} = S_{4}^{(b)} = r_{13} + r_{32} + r_{24} + r_{41}$ for all mass points.}
  \label{fig2}
\end{figure}
where $P_{4} = r_{12}r_{23}r_{34}r_{41}$ and $S_{4} = r_{12} + r_{23} + r_{34} + r_{41}$.  We see immediately from Fig.~\ref{fig2} that, generally, $S_{4}^{(a)} \neq S_{4}^{(b)} = r_{13} + r_{32} + r_{24} + r_{41}$, which represents a separate 4-body contribution to the self-energy.  It follows that even in a simple 4-body diagram one cannot ensure that $\cos(\mu S_{4}^{(a)})$, $\cos(\mu S_{4}^{(b)})$, $\sin(\mu S_{4}^{(a)})$, and $\sin(\mu S_{4}^{(b)})$ will all vanish for all mass points located at any $\vec{r}_{1}$, $\vec{r}_{2}$, $\vec{r}_{3}$, and $\vec{r}_{4}$.

Given the extremely large coefficients arising from Eqs.~(\ref{binomial})--(\ref{M U ratio}), we conclude that the cancellations suggested in Refs.~\cite{Stadnik} and \cite{Smirnov} are extremely unphysical as a mechanism for cancelling the many-body neutrino-exchange contributions to the self-energy of a neutron star or white dwarf.  As shown in Eq.~(5.39) of Ref.~\cite{Woodahl}, the 8-body interaction energy $W^{(8)}$, for example, is according to Eq.~(\ref{parameter}) proportional to $\eta(N,R_{10})$ and is given numerically by
\begin{equation}
W^{(8)} = 5 \times 10^{77}~{\rm eV} = 3 \times 10^{11} M_{1}c^{2},
\end{equation}
where $M_{1}$ is the mass of the Hulse-Taylor pulsar.  For the higher order contributions the combinatoric factors make such cancellations unphysical, as they would have to apply to each  of an infinite number of possible sub-volumes for $R \gtrsim 10^{-6}$~cm, as well as extremely large numbers of individual neutron stars and white dwarfs.

It follows that the conclusion of Ref.~\cite{Fischbach AoP} remains correct: For any neutrino species, $m_{\nu}\gtrsim 0.4~{\rm eV}/c^{2}$ is required to ensure that many-body neutrino-exchange forces do not lead to unphysical large many-body neutrino-exchange contributions to the self-energies of neutron stars or white dwarfs.

%\begin{acknowledgments}
%\end{acknowledgments}

\end{document}